\documentclass[11pt,a4paper,english]{article}

\usepackage{geometry}
\usepackage{graphicx}
\usepackage{fancyhdr} 
\usepackage{lastpage} 
\usepackage{amsmath,amsfonts,amsthm} 
\usepackage[svgnames]{xcolor} 
\usepackage{booktabs} 

\usepackage{authblk}
\usepackage{setspace}
\usepackage{hyperref}
\title{How to create an artificial magnetosphere for Mars}

\author{R.A. Bamford$^1$, B.J. Kellett$^1$,  
		J. Green$^2$, C. Dong$^3$,
		V. Airapetian$^4$, \\ 
		R.Bingham$^{1,5}$.}

\affil{$^1$RAL Space, STFC, Rutherford Appleton Laboratory, Chilton, Didcot, OX11 0QX, UK.\\
	$^2$NASA Headquarters, Washington DC, U.S.\\
	$^3$Department of Astrophysical Sciences, 4 Ivy Lane, Princeton University, Princeton, NJ 08544, U.S.\\
	$^4$NASA Goddard Space Flight Center (GSFC), 8800 Greenbelt Rd, Greenbelt, MD 20771, U.S. NASA ARC-SST, U.S.\\
	$^5$SUPRA, Dept of Physics, Uni of Strathclyde, Glasgow, Scotland, U.K.}

\usepackage[ddmmyyyy]{datetime}
\usepackage{datetime}
\newdate{date}{16}{09}{2021}
\date{\displaydate{date}}

\begin{document}

\maketitle

\begin{abstract}
If humanity is ever to consider substantial, long-term colonization of Mars, the resources needed are going to be extensive. For a long-term human presence on Mars to be established,  serious thought would need to be given to terraforming the planet. One major requirement for such terraforming is having the protection of a planetary magnetic field - which Mars currently does not have. The Earth’s magnetosphere helps protect the planet from the potential sterilizing effects of cosmic rays and also helps retain the atmosphere, which would otherwise by stripped by large solar storms as they pass over the planet. Mars does have small patches of remnant surface magnetic field, but these are localized in the southern hemisphere and are not of sufficient size or magnitude to protect the planet or a colony. 

In this article we explore comprehensively for the first time, the practical and engineering challenges that affect the feasibility of creating an artificial magnetic field capable of encompassing Mars. This includes the concerns that define the design, where to locate the magnetic field generator and possible construction strategies. The rationale here is not to justify the need for a planetary magnetosphere but to put figures on the practicalities so as to be able to weigh the pros and cons of the different engineering approaches. 

The optimum solution proposed is completely novel, although inspired by natural situations and fusion plasma techniques. The solution with the lowest power, assembly and mass is to create an artificial charged particle ring (similar in form to a ‘radiation belt’), around the planet possibly formed by ejecting matter from one of the moons of Mars (in fashion similar to that that forms the Io-Jupiter plasma torus), but using electromagnetic and plasma waves to drive a net current in the ring(s) that results in an overall magnetic field. 

With a new era of space exploration underway, this is the time to start thinking about these new and bold future concepts and to begin filling strategic knowledge gaps. Furthermore, the principles explored here are also applicable to smaller scale objects like manned spacecraft, space stations or moon bases, which would benefit from the creation of protective mini-magnetospheres.
\end{abstract}


\section{Introduction}

\subsection{Why do we need a magnetosphere for Mars}

The Earth’s magnetic field that originates within the iron core from a dynamo process, encompasses the planet and extends out into the near space environment (see Figure 1). The magnetic field helps to reduce the radiation reaching the surface by re-directing and shielding large numbers of energetic solar particles that would otherwise create a radiation hazard to life. Another important benefit of the Earth’s magnetic field is that it inhibits the loss of atmospheric molecules from pick-up by the solar wind during large solar superstorms \cite{[1.]}\cite{[2.]}\cite{[3.]}. Increasing Mars atmospheric pressure has been proposed as one of the primary requirements in terraforming Mars, along with warming and altering the atmospheric composition (e.g. \cite{[4.]}\cite{[5.]}\cite{[6.]}\cite{[7.]}\cite{[8.]}\cite{[9.]}\cite{[10.]}\cite{[11.]}). The aim is to achieve a stable ecosystem or ‘ecopoiesis’\cite{[12.]}\cite{[13.]}\cite{[14.]}. But recent studies suggest that these efforts would be undone by a combination of processes driven by extreme ultraviolet light and solar wind from the Sun, removing atmospheric gases from the upper atmosphere to space \cite{[15.]}\cite{[16.]}\cite{[17.]}\cite{[18.]}. Refs. \cite{[19.]}\cite{[21.]}\cite{[20.]} show that the presence of a strong intrinsic global magnetic field substantially decreases the loss of molecular ions and alters atmospheric conditions. 

In contrast smaller, sub-global magnetic fields offer a mixed benefit. The evidence from observations and simulations of the patches of crustal magnetic field that naturally occur already on Mars show that the presence of these anomalies can aid ion loss as much as they might hinder at other times depending upon the orientation of the field and interplanetary environment(e.g.\cite{[23.]}\cite{[24.]}\cite{[25.]}).

Mars is about half the size of the Earth and has a much lower atmosphere density. This therefore makes atmospheric losses much more significant. Terraforming activities designed to build up the atmospheric pressure and alter its composition on Mars will not want this effort to be undone by the first significant solar superstorm to reach the planet. One of the first goals of terraforming will be to increase the atmospheric pressure above the Armstrong Limit (6.3 kPa), a threshold that removes the requirements of having to wear a full-body pressure suit, although oxygen will still be needed \cite{[26.]}. Below the Armstrong atmospheric pressure limit, water in the lungs, eyes and saliva spontaneously boils \cite{[27.]}. Changing the atmospheric pressure can be expected to have wide ranging consequences to many aspects of living and working on Mars including amongst others to weather patterns, dust storms and transportation to name but a few. Primarily though, a global magnetic field generated magnetosphere, Mars could weather the worst of the atmospheric stripping effects of large solar events and help protect from radiation particles.

The past several years has seen an increase in the number of serious scientific investigations of many diverse aspects related to manned exploration of Mars and colonization. These include potential missions, interplanetary vehicles, Mars transportation vehicles, habitats but also socioeconomic concerns (e.g. \cite{[30.]a},\cite{[31.]a}\cite{[32.]a}\cite{[33.]a}\cite{[34.]a}\cite{[34.]b}\cite{[35.]a}\cite{[36.]a}\cite{[38.]b}\cite{[37.]a}\cite{[38.]a}). This indicates that the technology is becoming closer to achievable and affordable.

In this article we shall consider potential technological approaches to create an artificial magnetic field to protect Mars. We will not discuss the value or likelihood of humanity colonizing Mars, nor consider the relative merits or performance of magnetospheres, whether they are generated by magnetic fields or otherwise. Nor shall we present an analysis of the possible changes to Mars atmosphere with and without a planetary magnetic field. Such atmospheric modelling requires dedicated articles and will depend on the choice of location of the magnetic field source - for instance below or entirely above the planet’s atmosphere. 

What will be presented are multiple options for technology approaches and locations for the magnetic field generating infrastructure along with their pros and cons. The assumption is made here that there is a desire to create a magnetic field similar to that of a natural magnetized planet like the Earth and then follow how this could be done from a purely fundamental perspective. This issue, of creating an artificial structure at unprecedented scale, has not been considered in a peer-reviewed journal before. The calculations of power, resources and other relevant parameters are all deliberately made only to first order, as higher precision figures would be meaningless without a comparable level of precision for the engineering. This can be undertaken later. However, before any more detailed engineering design can be proposed there must first be an evaluation of the benefits and limitations of the different approaches and a choice of principle made. The aim here therefore is to discuss and compare the methods and to finally propose a novel solution.

The technological options we will consider include: re-starting the planet’s iron core, using solid state permanent magnets in either continuous loop or a series of discrete magnets, the use of solid state superconductors or a plasma current loop similar to a current driven plasma torus of an artificial plasmasphere. We shall also consider some of the factors concerning the source location of these generated magnetic fields. Within this analysis, we shall outline the issues and concerns that define the design such as general mass and electrical current needs. Specific timescales and logistics of installation will not be considered here, as it is anticipated that terraforming Mars will be a worldwide and multi-century endeavor and the potential for paradigm-changing developments would radically alter these. The one exception is the assumption of the development of successful nuclear fusion reactors \cite{[28.]}  as an efficient energy generation option. Nuclear fusion is already an extensive international scientific and engineering program that is ever closer to being achieved \cite{[28.]}. Fusion power is a likely necessary enabler for considering substantive colonization and terraforming in general. Fusion based propulsion has been proposed as an important development for human planetary exploration \cite{[29.]}, although at this time a successful economic fusion reactor has yet to be developed.

\begin{figure}
	\centering
	     \includegraphics[width=0.5\textwidth]{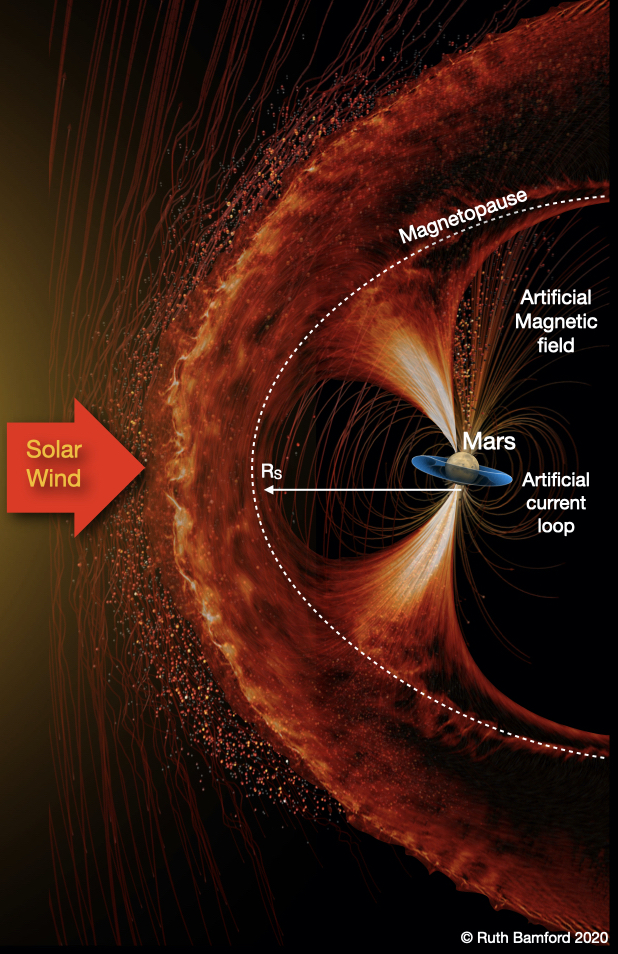}
	     \caption{An artistic impression of magnetosphere around Mars formed by a magnetic dipole field from an artificial ring or loop of electrical current circulating around the planet. The approximate point at which the pressures balance is the stand-off distance  $R_S$. }
	     \label{FIG:1}
\end{figure}	   

\section{Power needed to create a planetary sized magnetic field}

The primary technical challenge in creating a magnetosphere for Mars is not the strength of the magnetic field but the vast size of the magnetic field needed to encompass an object as large as a whole planet.

To first order, for a magnetic field in the path of a flowing plasma (like the solar wind), a stand-off will occur approximately where the magnetic field pressure $P_B$ balances the pressure of the solar wind plasma, $P_{sw}$\footnote{In practice there are number of other factors that refine the magnetopause stand-off distance such as the magnetosheath conditions, dipole tilt, interplanetary conditions etc 9.}. For planetary magnetospheres, this distance $R_s$ is sometimes called the Chapman-Ferraro distance in reference to those that first proposed it \cite{[30.]}.

\begin{equation}
	R_S^6\sim \frac{B_o^2}{2\mu_o P_{sw}}.R_0^6
\end{equation}\label{EQU:1}

Here $R_0$ is the radius of the magnetic field generating structure (iron core, or current/magnetic loop). 

For Earth, the magnetopause distance ranges from about 5 to 15 Earth radii depending upon conditions. For Mars, without an intrinsic magnetic field, an induced magnetosphere is created close to the planet or in the ionosphere \cite{[34.]}. For Mars this has been observed to mean that during solar superstorms there is a considerable loss of atmospheric molecules \cite{[1.]}\cite{[31.]}\cite{[32.]}. The creation of an artificial magnetic field will help limit this \cite{[19.]}. We can assume a minimum requirement for an artificial magnetic field for Mars will need to be such that, even during largest solar events, the $R_S$ never goes into the planet’s ionosphere.

The solar wind pressure is usually in the range $\sim$1 to 10 nPa ($1-10\times10^{-9}Nm^{-2}$) at Earth making the magnetic field intensity (BS) necessary to balance the solar wind ram pressure of $\sim$50-200 nT. This is using magnetic field pressure $P_B=B_S^2/2\mu_0$, where $\mu_0$ is the permeability of free-space. At the orbit of Mars this would be $\sim$40-150nT.

From this (as observational data from spacecraft that confirm \cite{[34.]}) we can see that the intensity of magnetic field at the magnetopause needed to disrupt the solar wind, is not very much less than the field of a typical fridge magnet at $\sim$ 5mT \cite{[35.]}. However, to cover an area at least the size of Mars would mean a $\sim$100nT magnetic field over a minimum of 37 million square kilometres (the radius of Mars is $R_M\sim3400$km + $\sim$100km atmosphere). The total energy stored in such a magnetic field is of the order of $10^{17}$J. This does not include the energy needed to ramp the magnetic field up, which will be affected by the inductive properties of the media in and around the magnetic field generating structure. This provides an absolute minimum for the energy needed as that stored in the minimal field once operational.

This amount of energy can be compared to the world’s total electricity consumption in 2020, which amounted to approximately 583.9 exajoules ($10^{18}$ joules) of energy or 553 Quadrillion BTU \cite{[36.]}. For this reason there would be a strong need to have working fusion reactors as an efficient, compact power source as a necessary precursor for embarking on the creation of an artificial magnetosphere for Mars and a general enabler for economic permanent colonization. Nuclear fusion as a power source has the highest energy density – making it compact and reducing the mass needed, which is an important consideration for transportation to Mars. The energy density of fusion fuels, at about $10^{12}$Jg$^{-1}$ of deuterium, is orders of magnitude higher than nuclear fission (next nearest) at $10^{10}$Jg$^{-1}$ for $U^{235}$ and without the radioactive hazard concerns. Nuclear fission also has an energy density millions of times higher than solar power (at 1 AU), or chemical sources like fossil fuel, biofuel, or batteries at less than $10^4$Jg$^{-1}$ \cite{[37.]}.

   \section{Approaches and Locations}
   
\begin{figure}
	\centering
	     \includegraphics[width=1.0\textwidth]{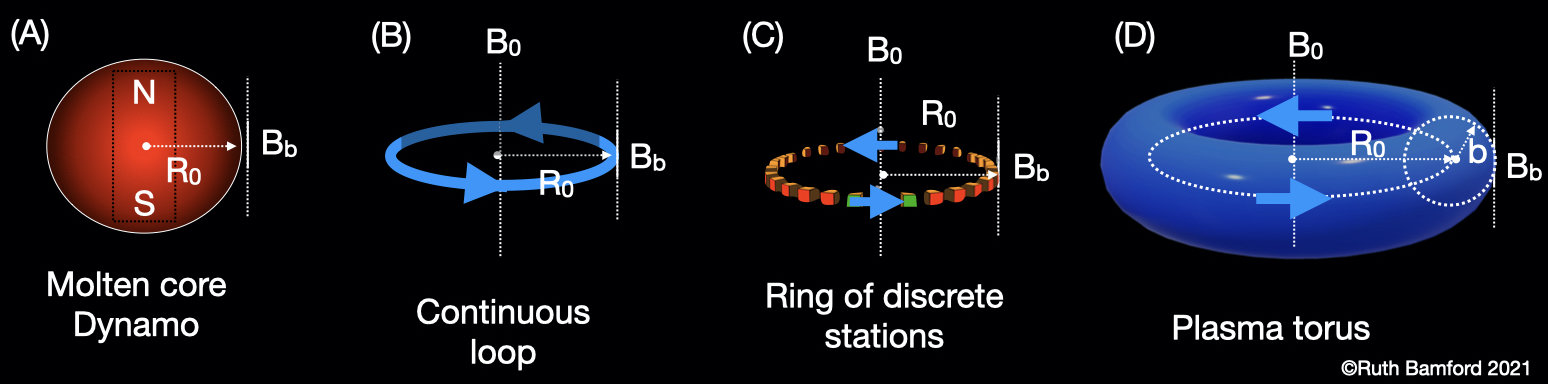}
	     \caption{ Approaches to creating a magnetic field. The options for the different approaches to creating a current loop are; (A) molten iron core dynamo, (B) solid superconducting current loop or permanent magnets, (C) a chain of discrete coupled current or magnetic sources and (D) a current driven plasma torus.}
	     \label{FIG:2}
\end{figure}
\begin{figure}	     
	     \includegraphics[width=1.0\textwidth]{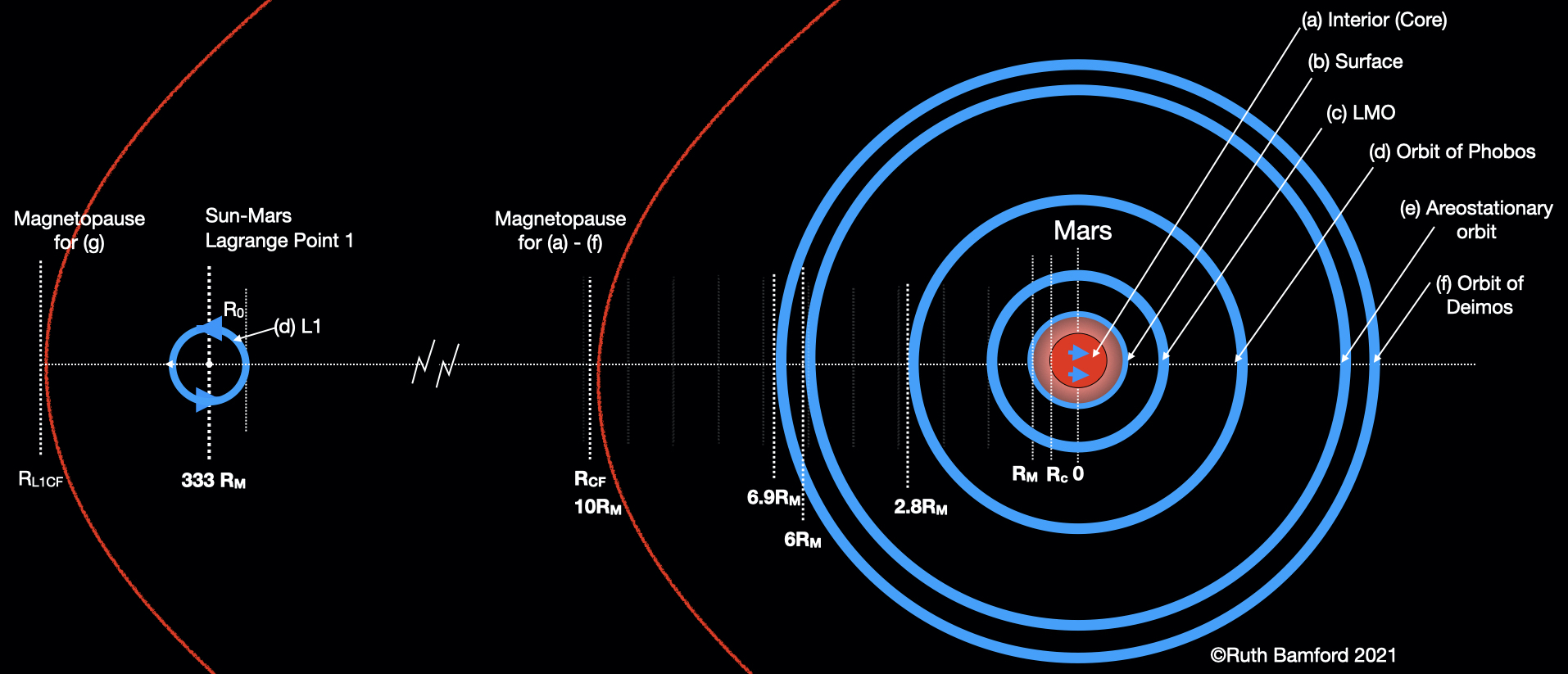}
	     \caption{ Options for locating a source of magnetic field; (a) from the core of the planet, (b) the surface of Mars, (c) in orbit of Mars (Low Mars Orbit (LMO) or areostationary orbit), (d) the orbit of Phobos, (e) the orbit of Deimos, and (f) the Mars-Sun L1 Lagrange point.}
	     \label{FIG:3}
\end{figure}

To form an artificial magnetosphere a magnetic field needs to be created artificially. There are several ways this might be done in principle. Figure 2 shows the options for the different approaches to creating a suitable current loop. The options are: 

\begin{itemize}
	\item[(A)] dynamo circulation of a molten planetary core, 
	\item[(B)] a continuous solid superconducting current loop or loop of permanent solid-state magnets, 
	\item[(C)] a chain of discrete coupled current sources made of a controlled beam of charged particles forming an electrical current, and 
	\item[(D)] a plasma torus of positively and negatively charged particles with artificial current drive forming a resultant current loop of a solenoid. 
\end{itemize}

At some distance from the source of the magnetic field, these solutions will all be indistinguishable from each other. The options for locating the source of magnetic field include in and on the planetary surface and a variety of locations in Martian orbit.

In the companion Figure 3 are shown the options for locating the source of magnetic field in 

\begin{itemize}
	\item[(a)] the planet’s core, 
	\item[(b)] on the surface, 
	\item[(c)-(f)] a variety of locations in Mars orbit, and 
	\item[(g)] at the Sun-Mars L1 point upstream of the planet (first proposed by J. Green) \cite{[21.]}.
\end{itemize}

We shall now consider the factors and practicalities of these options in turn, starting with what would be required to re-start Mars’ iron core with the intention of activating a dynamo in the same manner as the Earth’s core generates its magnetic field.

\subsection{Restarting Mars’ iron core}

The Earth’s magnetic field originates from dynamo effects in the outer iron core kept molten in part due to radioactive material and pressure \cite{[38.]}, \cite{[39.]}. Earth’s iron core is approximately 3,500 km in radius (by coincidence very similar to the size of the entire planet of Mars \cite{[40.]}.) Mars’ iron core is small at approximately 1800 km (see \cite{[41.]},\cite{[42.]}, \cite{[43.]} for recent reviews). Recent observations by the NASA InSight mission have suggested that the core is in a liquid state \cite{[44.]}. Therefore, there exists the possibility that one could kick-start Mars’ iron core back into an active magnetic dynamo, possibly using nuclear material or electrical (inductive) heating. The amount of energy needed, ignoring all other issues such as drilling, would be of the order of $10^{26}$ Joules to raise the $\sim2\times10^{16}$ m$^3$ volume of the iron (II) sulphide \cite{[45.]} core by 1000-2000K (assuming a mean core specific heat of $\sim800$Jkg$^{-1}$ C$^{-1}$ and density of $8\times10^3$ kg m$^{-3}$). Since a 1 megaton Hydrogen bomb has an equivalent energy release of $10^9$ kg of TNT, which is an amount of energy equal to $4\times10^{15}$ J \cite{[46.]}, the amount of energy alone needed to melt Mars’ iron core is the equivalent of $10^{11}$, 1 megaton H-bombs. This ignores any other issues with the placement and timing of explosions and any subsequent heat losses. However once molten, the retained energy would provide the heat-engine and reduce the viscosity of the core material that would allow the Coriolis forces to create the convection currents that could self-excite the magnetic dynamo into producing a resultant magnetic field. Although the compositions and pressures of Earth and Mars cores may be very different, using a simple proportionate core and planet size argument between Earth and Mars, a restarted core circulation would provide an estimated magnetic field intensity of approximately $10^{-4}$ T at the surface of Mars.

Even so it is not known how long such a magnetic field would last, given that it is believed early Mars did possess a magnetic field that it lost \cite{[41.]}. In addition, Mars’ mantle surrounding the core, is about half the thickness of Earth’s mantle making the heat flow and potential uncertainty over the tectonic stability at the surface more significant with a restarted core with convective motions.

The volume involved is vast and it is highly unlikely that this option would be feasible. It is also unnecessary since there are other potentially much easier alternative ways of creating an artificial magnetic field that may be more practical. So let us now consider the physics behind solid magnets and current loops as potential sources of magnetic field for creating a magnetosphere \cite{[47.]}.

\subsection{Solenoid loop}

If $R_0$ is the major radius of the current loop or coil with a total current of I (Amps) where $I=NI_N$, where $N$ is the number of current loops or sub-coils each carrying a current of $I_N$, the following approximations can be applied (in SI units), for a solenoid, the field at the center of the coil is:

\begin{equation}
	B_o\simeq \frac{\mu_0 I}{2R_0}
\end{equation}\label{EQU:2}

Clearly a permanent magnet approach does not involve currents but the expression for the decrease of the magnetic field intensity with distance will follow the same equation below.
At a distance $R\gg R_0$ (in any direction) the decrease in magnetic field intensity $B(R)$ from a dipole is;

\begin{equation}
	B(R)\simeq B_0 \left (\frac{R_0}{R} \right )^3
\end{equation}\label{EQU:3}

Here $R_0$ is the major radius of the magnet or current loop. Magnetic field configurations with higher order poles decrease with distance more rapidly closer to the source. Far from the magnetic field source, all magnetic field configurations reduce to a dipole minimum form because of $\nabla\cdot B=0$. We assume for simplicity that the magnetic field within the solenoid is uniform. (In fact with a wide major radius the magnetic field could decrease within the loop depending upon the configuration of neighboring coils and minor radius of current loop.)

What Equation 4 shows us is that the magnetic field decrease with distance is not just $\propto 1/R^3$ but $\propto  (R_0/R)^3$, and just $\propto B_0$, meaning $R_0$ is a much more valuable parameter than $B_0$ and the optimum stand-off distance is achieved with the largest possible coil major radius, $R_0$ \cite{[47.]}. Furthermore, the gradient of the decrease is less\footnote{ In the very far field when $R\gg R_0$ this will cease to be significant.}.

To protect the entire planet, the magnetic field must also cover all latitudes. This means the current loop would be required to either have a wide minor radius (panel (a) in Figure 4) or to be made up of narrow minor radius coils in a Helmholtz (or similar) configuration (for instance) as in panel (b) of Figure 4. This would produce a continuous magnetic dipole field indistinguishable at a distance from that of a natural field. The plane of the coils does not necessarily have to be that of the ecliptic. 

 The configurations illustrated in Figure 4 show the target area of Mars as seen from the perspective of the Sun and the solar wind and current loops and resultant total magnetic field located surrounding and close to the planet. The same consideration for off-major axis plane magnetic field characteristics is needed regardless of location or orientation.

\begin{figure}
	\centering
	     \includegraphics[width=1.0\textwidth]{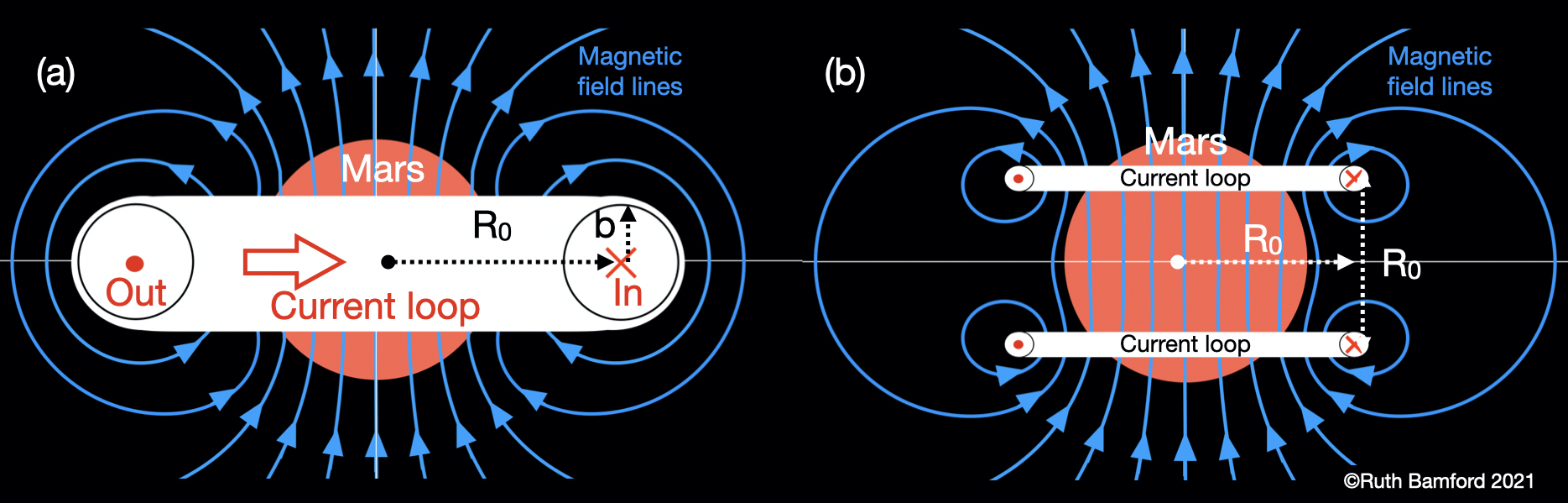}
	     \caption{ To cover the entire area of the planet would require either (a) A wide minor radius, b, current loop (which can be made up of N current loops) or (b) a set of more discrete coils/current loops in Helmholtz (or similar) configuration.}
	     \label{FIG:4}
\end{figure}

As well as the greater range of wider major diameter, low field coils compared to a smaller high field coils, another potential advantage is that it provides for a potentially safer magnetic field intensities for humans and instrumentation for those working and living close to the source of the magnetic field. It reduces the risk of strong gradients in magnetic field intensity – potentially a more significant hazard than a constant high field. Strong gradients in magnetic field and rapid changes in magnetic field in time, (such as experienced by a spacecraft passing the field coils) could induce large electric fields and currents in instrumentation and humans alike. For a planetary wide artificial field, the ability to work on and around the structure will be very important. Having a magnetic field intensity of the same order as that on the surface of the Earth would clearly not pose a hazard. Most of this argument will also hold for a permanent magnet in the same form of a loop. Once outside the material creating the magnetic field, there is no physical difference. The draw back of a wide diameter current loop is the scale of the structure. This will be discussed later.

\section{Coil locations options: surface, orbit or L1}

Some optimizations between resources in construction and maintenance with operational needs would determine the optimum location. Low Mars Orbit (LMO) would have the advantage of being the easiest to reach although LMO is subject to drag from the planet’s tenuous outer atmosphere. The atmosphere of Mars is significantly thinner than that of Earth, with a surface pressure of just 1\% of the Earth’s. However, because of the much smaller size and mass of Mars, the two atmospheres have a similar scale height. So, objects in LMO would still require periodically boosting in order to avoid orbital decay and re-entry. Here we want to consider some of the specifics.

Taking the example of Earth’s equatorial surface magnetic field strength of $B_0 =31,100$nT and radius of Earth $R_E=R_0=6,400$km and using equation 2 above, requires a current $I=3\times10^8$Amps or 0.3 GAmps. However, for Mars the size of the planet is a lot less ($R_M=$3,400km) so that the magnetic field need not be that of the surface of Earth. 

The necessary criteria are:
\begin{enumerate}
	\item The magnetic field needs to be of sufficient intensity to create an artificial magnetopause.
	\item The size of the protected zone of the magnetosphere needs to encompass the entire planet.
	\item The magnetosphere must not enter the planet’s atmosphere during extreme space weather events (thought to be the major source of atmospheric loss).
\end{enumerate}

Determining the parameter ranges for dynamic pressure balance, if the normal solar wind \cite{[48.]} ram pressure is taken as ranging between 3 and 10nPa, a severe solar storm flux of $10^{10}$ protons/cm$^2$sec at $>$50MeV would correspond to around 8000nPa (if treated as $\rho v^2/2$ (deflected flow) or 16000nPa if treated as $\rho v^2$ (absorbed flow)). For normal solar wind conditions, a field of 90-150nT is sufficient to achieve stand-off balance (see equation 2). As has already been mentioned, the magnetic field at the magnetopause of the Earth is of the order of $\sim100$nT. If we choose to have the magnetopause stand-off distance to be a similar number of planetary radii as the Earth’s, then $R_S\sim10R_M$, where $R_M=3,400$km is the radius of Mars. This is relatively arbitrary but allows for plenty of margin in the solar storm extremes.

The question then becomes where to locate the solenoid loop, bearing in mind the analysis above. For simplicity here we shall illustrate using single generic ‘current loops’ while acknowledging the concerns discussed in the previous section. The options shown in Figures 3 are (A) the iron core (covered already), and current loops locations - (B) on the planetary surface, (C) to (F) in orbit, and (G) located at L1. On the surface of Mars, the current needed would be $>$5GAmps producing a surface magnetic field of $\sim10^{-4}$T - approximately twice the magnetic field at the surface of the Earth and well within the safe limit for any humans to live with. However, there may well be issues about living and working around the required structures on the surface. Spatial and temporal gradients in magnetic field induce electric fields and are a greater hazard than constant magnitude of the magnetic field. This may be a problem if we consider some of the practical issues with managing a large single loop structure. The facility needs to be able to handle the energy dissipated if the coil is quenched (assuming superconductive coil). The quenched current could be directed to a reversed induction non-superconducting conduit, such as a copper jacket around the coil. Another option is to have a series of connections rather than a single continuous loop, where there are joints or release points. This may have mechanical or energy efficiency drawbacks, but is probably still necessary to allow for maintenance. Some of the issues of living and working near the structure would be avoided if the structure were in orbit. This also allows for a larger $R_0$ and correspondingly lower $B_0$.

\section{Solenoid locations in space}

\begin{figure}
	\centering
	     \includegraphics[width=0.5\textwidth]{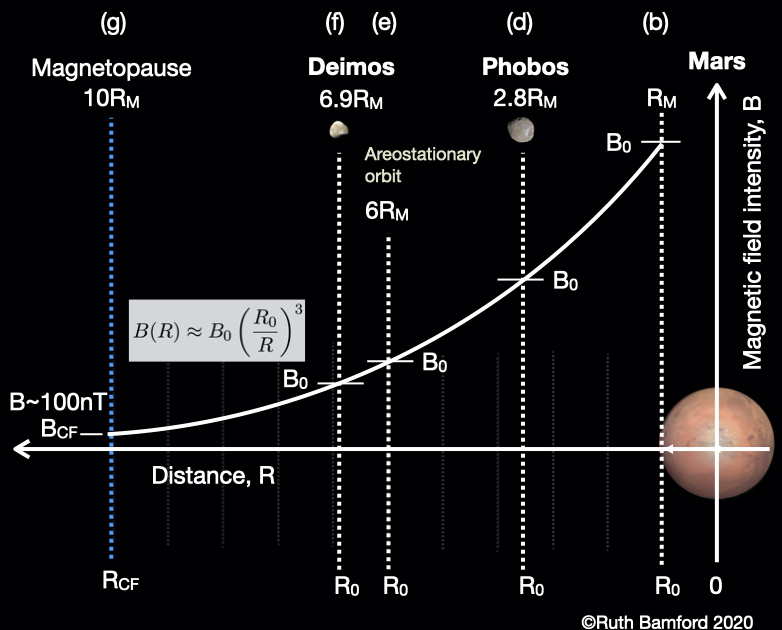}
	     \caption{  The options for locations of a current loop in orbit include Low Mars Orbit, the orbits of the Moons (Phobos and Deimos) and the Mars equivalent of geostationary, areostationary orbit. If we fix the location of the magnetopause at $10R_M$ with a field of 100nT then the magnetic field intensities created must follow the curve shown and the wider the loop the lower the field, but the larger the structure needed.}
	     \label{FIG:5}
\end{figure}

Placing a solenoid in orbit requires a stable orbit. Mars has a very uneven gravitational pull due to large asymmetries in mass, making station keeping particularly problematic, especially for LMO. This would require considerable delta-V to maintain the orbit \cite{[50.]}.

If a solenoid were to be placed in an areostationary orbit for Mars (the equivalent of Earth’s geostationary) of a $R_0=6R_M$ then the much wider loop radius drastically reduces the magnetic field intensity needed to reach $\sim$100nT at $R_S$=10$R_M$ of 460nT and consequently requires more than an order of magnitude less current at $<0.2$GAmps. This is illustrated in Figure 5. During extreme superstorms the magnetopause will move from 10$R_M$ to closer to less than 3$R_M$ (assuming the storm values given previously in Section 2 and reference \cite{[21.]}). We assume that this is still sufficient to still protect the planet\footnote{A more detailed analysis would be needed to determine if this is indeed sufficient distance.}. It would be sensible to build in the ability to increase this margin by increasing the current in the coils when needed, and therefore increase the magnitude of the magnetic field for the relatively brief time of the duration of an extreme solar event – which can last from hours to a few days.

Figure 4 shows the possible locations for the coil (b) on Mars’ surface, (c) in an unspecified LMO, or (d) at the orbit of moon Phobos, (e) in areostationary orbit, (f) in the orbit of moon Deimos, and (g) smaller diameter coil at the Sun-Mars Lagrange L1 point. Creating an artificial magnetosphere at the L1 point \cite{[21.]}, makes use of the gravity balance point between Mars and the Sun to aid station-keeping of the hardware. The Mars L1 is 333 $R_M$ away and the question is then whether Mars would reside within the magnetotail created by the coil. This would depend upon the size of the magnetosphere allowing for the balance of pressures radially from the Sun and those following the Parker Spiral \cite{[51.]} which is more acute at Mars at $\sim20^\circ \pm 10^\circ$ . The benefit of L1 location is the coil can potentially be smaller and the magnetic field larger without concern for the magnetic field intensity being a hazard to colonization and operation on the planet due to the distance. However, magnetotails are known to undulate and move around with the variations of the solar wind parameters. This may make maintaining the planet in a ‘safe zone’ problematic, especially given the distance of the the Mars L1 point from Mars. An analysis of the performance of a magnetic field located at Mars L1 is presented in ref. \cite{[21.]}.

Station keeping at the Mars L1 point is more challenging than that for Earth. Mars L1 is a very shallow gravitational ``island'' and Mars’ orbit around the Sun is significantly non-circular so the Sun-Mars L1 point will move by a considerable distance during the 687 days Martian year. Nevertheless, a high field, narrow diameter multiple coil system at Mars L1 would offer a significantly lower mass option with greater safety margins that would more than make up for the extra delta-V needed.

If the current ring were to be made of solid-state materials, either superconductors or permanent magnets, and assuming the same minor cross section, then the mass budget would be a factor of 6 greater at the areostationary orbit compared to that for the surface. This brings us to the issue of mass.

\section{The problem with mass}


\subsection{Permanent magnets and superconductors}

Many of the difficult technological issues relate to the need to create a magnetic field using either permanent magnets or from an electromagnet (solenoid) that only acts like a magnet when an electric current is passing through it. As described above, the factors for either solution are essentially the same, because we are considering such modest field strengths. The major issue is the trade-off between magnetic field intensity at the source and diameter of the source structure.

It is reasonable to expect that, by the time that it becomes practical to terraform Mars, considerable effort and development will have been made that will have led to superconducting materials that operate at much higher temperatures even than liquid nitrogen (current technology) since the current research in the field is heading in this direction \cite{[52.]}\cite{[53.]}.

If the current carrying conductors are to be solid state, their material will need to have a very low resistivity so as to reduce the energy being lost in the forms of heat within the conductors. High temperature superconductors, that essentially have no electrical resistance would be the ideal material.


Creating permanent magnetic structures of the scale required would be very challenging. Current ferromagnetic and ferrimagnetic materials on Earth are based on iron and rare-earth elements and these are relatively high density so have high resultant mass. There is potentially an alternative that would avoid or reduce this issue – ferromagnetism has been demonstrated in a lithium gas cooled to less than one kelvin \cite{[59.]}.

Comparing this with superconducting solenoid technology currently these are either pure metals and metalloids that experience a sudden decrease in resistance with low temperature, or are metallic compounds or alloys, using elemental vanadium, technetium, and niobium. At this time, superconductors need to be cooled to at least liquid nitrogen temperatures to exhibit superconductivity. However, it is fair to assume that future superconductors will operate at higher temperatures than this, given the way the technology is progressing \cite{[60.]}.

\subsection{Hollow solenoids}

Due to the size of the structures being discussed here we are inevitably considering hollow loop structures. This radically effects the mass budget. The hollow solenoid coils could be made lattices such as is shown in Figure. 6 which in turn could be constructed of high-performance light weight superconducting (ideally) materials such as nanotubes. Carbon nanotubes (CNT) \cite{[54.]} exists today although they are not superconducting. Any such structure would have to hold the required current density without melting if there is finite resistivity. The electrical conductivity of CNT is $\sim10^6-10^7$ S/m and $10^8$ S/m for pure graphene \cite{[55.]}. If we take the solenoid current calculations from above section, which range between $\sim$0.2-0.5 GAmps (depending upon location) then this would require power of $\sim10^8$ to $10^{11}$ W using CNT. While there has yet to be built a fusion-based power station a 2 GW Demonstration Power Plant, known as Demo, is expected to demonstrate large-scale production of electrical power on a continual basis by around $\sim$2040 \cite{[28.]}. This is in the middle of the power range needed. To avoid melting the CNT which would start to occur at about $\sim$2600K \cite{[56.]}, using a thermal conductivity of $\sim$2600 -4000 W/mK \cite{[57.]} then the  total area of CNT needed would be of the order of $10^{11}$ m requiring a mass of $\sim10^7$kg assuming a $\sim$100nm depth of density of $\sim$1.4g/cm$-3$\cite{[56.]}. While this is not an astronomical quantity of mass and power, the required engineering infrastructures to mechanically support and power such a current loop would need to accommodate contingencies, and backups so a more realistic logistical burden estimate is likely to be far greater. 

\begin{figure}
	\centering
	     \includegraphics[width=0.5\textwidth]{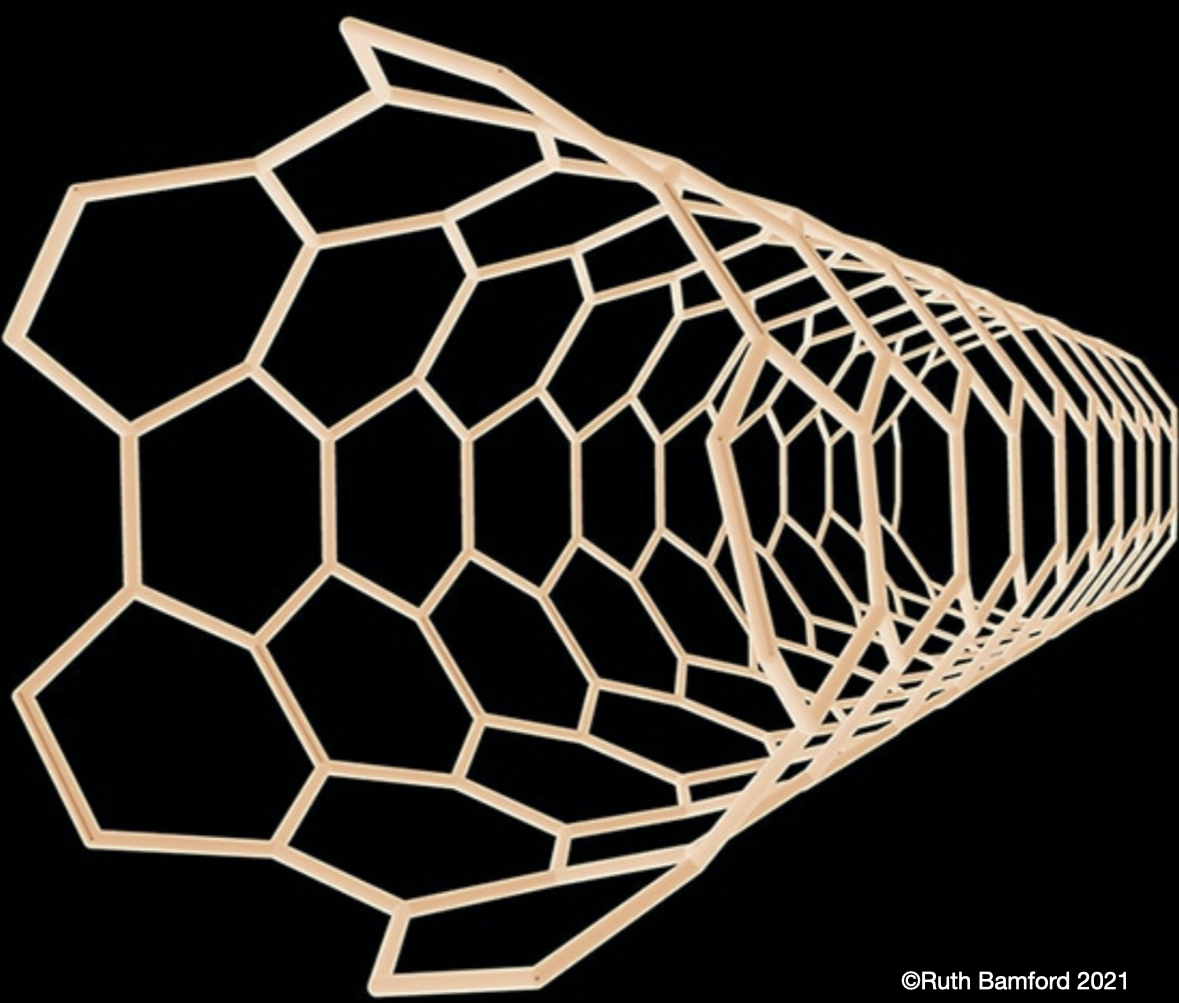}
	     \caption{Solenoid coils could be made from hollow lattice structures to minimize mass. These structure in turn could be comprised of lightweight materials such as nanotube materials.}
	     \label{FIG:6}
\end{figure}


One encouraging recent development is that of high-quality ultra-thin superconducting films \cite{[58.]}. This technology could be a game changer for using magnetic fields in space as it has low mass and seems robust enough to deal with launch stress and thermal environment of space. In an era of terraforming Mars, development of this type of material could be a significant driver for such innovations.

Many of the raw materials could potentially be found on Mars itself – they will obviously need to be refined and processed (using significant energy resources) and then the material will need to be raised into the required orbit. They will also need to be powered or cooled (or both) in order to operate. The gravity well of Mars being 38\% that of Earth will make raising the material to orbit somewhat easier but nevertheless this would require significant fuel, and this makes mass a critical consideration. The exact amount of mass needed to be used in orbit clearly depends very much on the technology used. Some of the options are considered in following sections.

For practical operations, each coil would need to be divided into two adjacent but distinctly separate windings so that any fault arising in one of them would not affect the other. This would also allow for down time and servicing.

\section{Placing structures in orbit}

For any orbital structure there is the problem of raising the materials into orbit (even assuming that the raw materials come from Mars). Figure 7 shows the relative gravity well for Earth, Mars and the Moon \cite{[61.]}. As can be seen Mars’ lower gravity makes lifting either materials or manufactured structures into space considerably easier than on Earth.

The escape velocity for Mars is less than half that of the Earth at 5 m/s. The lower air density on Mars will also make this easier. Mars’ moons, Phobos and Deimos are too small to be visible on the figure. Phobos and Deimos offer an obvious resource for the materials needed to construct an artificial magnetic field source in orbit. However, for the size of structures discussed above the moons would not be of sufficient mass. To determine the suitability of these and Mars for resources requires detailed knowledge of the mineral resources and abundances.

\begin{figure}
	\centering
	     \includegraphics[width=1.0\textwidth]{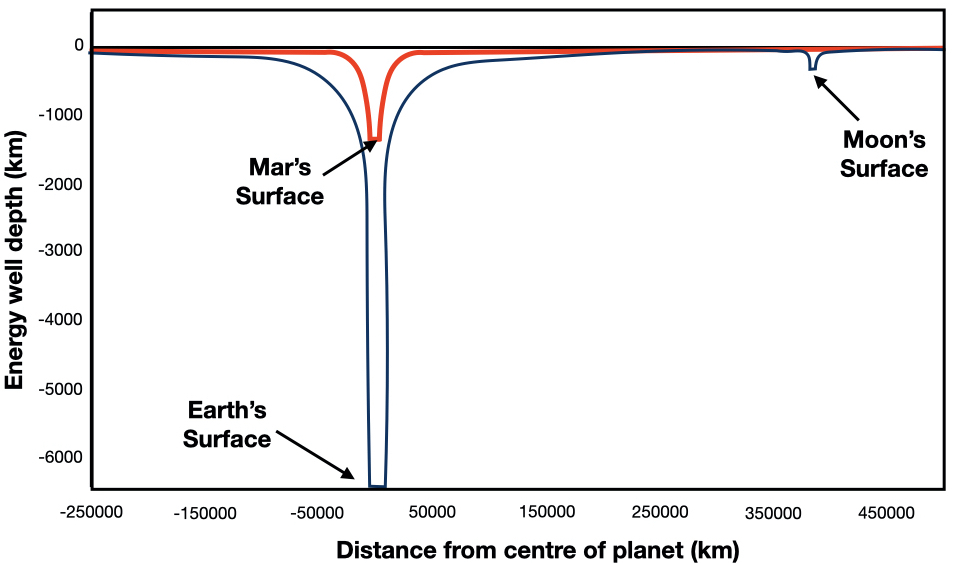}
	     \caption{ A comparison of the relative gravity wells between Earth, Mars and the Moon. The gravitational potential is expressed in units of kilometers and is the equivalent distance needed to raise a 1 kg mass in a uniform 1 g (9.8ms$^{-2)}$) gravity field. Adapted from Crawford (2015) \cite{[61.]}}
	     \label{FIG:7}
\end{figure}
 
It is likely the relative Martian abundances of the resources needed to construct an artificial magnetic field structure will play a large part in the type of technology employed. There is no point considering planetary scale structures that would need rare and exotic minerals in large quantities. Currently, the mineral resources of Mars are not well known or quantified. Iron is likely to be plentiful, particularly in the region around Olympus Mons and the other three giant shield volcanoes in the same region \cite{[62.]}. The mineral composition of Mars’ moons is even less well known. However, Phobos and Deimos are believed to be small captured asteroids at 22.2 km and 12.6 km in diameter with total masses of and $1.1\times10^{16}$kg and $1.0\times10^{15}$kg respectfully \cite{[63.]}. However, their example suggests materials from captured asteroids may be an option.

\section{Plasma torus}

\begin{figure}
	\centering
	     \includegraphics[width=1.0\textwidth]{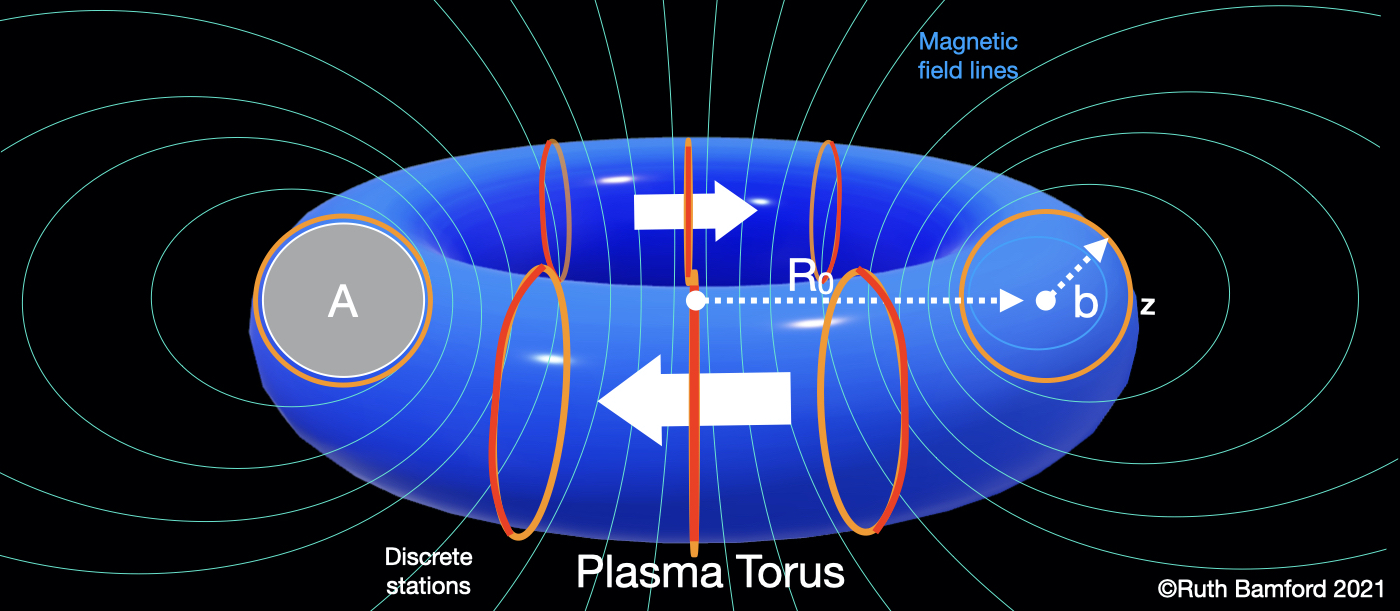}
	     \caption{ The principle of a plasma torus with current drive that produces a resultant magnetic field. Charged particles are directed between a series of space stations that guide the particles to form the current loop.}
	     \label{FIG:8}
\end{figure}

There is one final alternative to creating a large scale, space-based high current loop that does not require creating a physical structure the dimensions of Mars, that is superconducting, but will not melt. This is to use a plasma torus with a resultant ring current necessary to create a resultant magnetic field. This is illustrated in Figure 8. The concept would be similar to having an open particle accelerator like the Large Hadron Collider (LHC) at CERN \cite{[64.]} but in space and without the goal of accelerating particles. Enrico Fermi first suggested putting a particle accelerator in Earth orbit to deal with the ever-increasing dimensions needed to reach the ever increasing energy of particles needed for fundamental physics research \cite{[65.]}. The problem with doing this for particle physics is that while the vacuum of space is conductively cold, it is radiatively very hot. This, however, would be an advantage to the plasma torus as the radiative energy could be used to produce the ionization.

In the ultra-high vacuum of space charged particles (electrons and ions) could be accelerated and “beamed” between discrete space stations to create a resultant toroidal current loop. Multiple space stations would be needed to turn the particle streams into a loop and to ensure that the electrical circuit could be completed, to prevent the space stations continuously becoming electrically charged. The stations would introduce the curvature as well as accelerate the particles and maintain the cohesiveness of particle bunches. (This is similar to the action of the particle “kickers” in terrestrial particle accelerator facilities \cite{[66.]}.)

The major advantage of this approach is the reduced mass needed rather than any planet encompassing structure solid structure. Plasma structures such as radiation belts naturally occur around planets like the Earth. In these cases, the co-rotating ions and electrons are formed as a result of the rotation of the planet and complex interactions of its natural magnetic field. Here we do the opposite, artificially driving a current in a plasma torus to create a resultant magnetic field. The current can be carried by a current loop where the charge carriers are distributed over an area A, where the total current could be made from a a sum of N beamlets of current $I_c$, where:

\begin{equation}
	I=NI_n=nAvQ
\end{equation}\label{EQU:4}

Here $A$ is the cross-sectional area of the plasma torus and $n$ is the number of charge carriers per unit volume (or charge density), $Q$ is the electronic charge of the particles and $v$ is the drift velocity of the charge carriers.

For current drive \cite{[67.]}, the toroidal momentum is transmitted to either the plasma electrons or ions. The velocity of the particles in the toroidal direction must exceed their thermal velocity. The electrons need to be given sufficient momentum so that the high inertia ions cannot create a charge balance that cancels the electron current. The ions too could be given an artificial velocity in the opposite direction to enhance the current. The current would then be  $I=AQ(n_iv_i – n_ev_e)$ where $n_i$ and $n_e$ are the number density of ions and electrons respectively and $v_i$, $v_e$ their directed velocities.

The principles of current drive in laboratory fusion plasma confinement devices, such as tokamaks, are well established. Clearly for this unique application new technology would need to be applied but for a review of the underlying principles of current drive in plasmas see \cite{[67.]}. There are a number of methods which utilize particle beams or radio frequency waves in any of several frequency regimes. Traveling waves may be induced in the plasma that accelerate the particles via phased arrays or coil arrays and use could be made of plasma waves or electromagnetic waves which exist in nature. The types of wave are varied, the optimum waves could be resonant with lower hybrid waves, Alfv\'en, ion cyclotron or electron cyclotron resonant or use more sophisticated modern ’wakefield’ approaches \cite{[69.]}. Further discussion of which approach is most suitable would be better suited to a stand-alone article.

In general terms in this case, the magnetic field at the outer surface of the plasma loop $B_b$, with minor radius of the plasma torus of b, is approximately,
\begin{equation}
	B_b\simeq \frac{\mu_0 I}{2\pi b}
\end{equation}\label{EQU:5}

If we assume the magnetic field within the solenoid is uniform then $B_b=B_0$ as used previously.

For a plasma torus solenoid by equating equations (2) and (6) with $B_0\sim B(b)$ this provides $b\sim R_0/3$ then $b\sim2000$ km the area $A=\pi b^2\sim10^{13}$ m$^2$ and if we assume $Q=1$ and the charge carriers are electrons then current density $j=nv=5\times10^{-5}$ Amp m$^{-2}$ for the plasma torus. The attraction of this system, as shown by equation (6), is the $1/R$ scaling of the resulting toroidal magnetic field, but the disadvantages would include the continuous power required to drive the current. Having a plasma current not in a solid conductor but exposed in space solves the mass problem but comes with its own technical challenges which we will now explore. 

Over 70 years of laboratory magnetic confinement fusion research have greatly increased our knowledge of how to confine and control plasma tori \cite{[72.]}. While the criteria for fusion in the laboratory are more exacting than required here, the problems of scale and location do present their own challenges. The stability of a current driven in a laboratory plasma torus is aided by imposing a guiding magnetic field in the same direction as the plasma current (toroidal or torus major axis). This plasma current results in the creation of the magnetic field in the poloidal direction (minor radius cross section). Close to the plasma column, this poloidal component of field aids confinement, using the ‘pinch’ effect that helps inhibit (though not eliminate) the radial expansion of the plasma particles. Different types of laboratory devices have different relationships between these three basic components of plasma current, toroidal and poloidal magnetic field \cite{[68.]}. Unlike in a natural environment, in an artificial system these are all variables available to be optimized. Here the aim is to maintain the integrity of the plasma current loop sufficiently to result in an overall dipole magnetic field and maintain it against the processes such as diffusion, instabilities, particle pick up and shielding. There are several ways to do this.

One way to prevent particles of the solar wind plasma from forming a responsive shell that cancels out the net magnetic field produced by the particle beam current, is to increase the velocity of the particles forming the current given by equation (4) so that their velocity is very much greater than the thermal velocity, $v \gg v_{th}$. Particles (generally electrons) like this in tokamaks are called ‘runaways’ \cite{[70.]}, \cite{[71.]} because they are essentially unmagnetized and too energetic to be effected by the bulk plasma flow and hence pass straight through it. In a similar way, highly energetic protons from some solar eruptions can cross the solar system, passing through all the regions of the Earth’s magnetosphere and reaching the ground \cite{[73.]}.  For this application creating such a directed beam would be an advantage as it would maintain the current and mean that the plasma particles carrying the current would not be easily ‘picked-up’ or disrupted by the external plasma flows from the space plasma environment, such as effects due to solar storms. To do this the velocity in equation (4) would need to be such that $v_e\sim10v_{th}\sim100$keV or $\sim0.1c$ where $c$ is the speed of light.

However, at runaway velocities, the plasma torus would be difficult to align and it would thus be difficult maintain a loop, forming a radiation hazard to any transiting spacecraft e.g., \cite{[74.]}. Therefore, an alternatively higher density, lower velocity scheme could be used. Here, since 1 ampere $=6.2\times10^{18}$ particles per square meter per second, if we pick a velocity of 1 $ms^{-1}$ (non-hazardous) the required density $n =3\times10^{14}$ charged particles per m$^3$ this density is so low as to not be a concern to spacecraft. Here the problem will be a greater diffusion of the particles leaking away. To  mitigate this, the stations would need to create a linked magnetic geometry or toroidal guide field. This could be done with an array of poloidal solenoids arranged in a toroidal loop. In laboratory plasmas, this configuration is historically known as a ``Polytron'' \cite{[75.]} or “bumpy torus” \cite{[76.]}. In such a configuration, the superposition of high-ripple toroidal magnetic fields acts as a guide field for the toroidal plasma discharge.

Once the magnetic field is created, the full force of the interplanetary solar plasma should not directly reach the current loop, as it will be inside the magnetosphere created. The evidence from natural magnetospheres is that even the non-current driven, natural plasma ‘torus-like’ structures such as radiation belts and plasmaspheres persist and are not regularly eroded by pick-up processes during storms, at least not significantly enough to be ‘blown away’. Sufficient density of higher Z materials will help retain the plasma. At times additional mass loading of the current loop may be needed to ensure dominance of the artificial current during extreme solar events. 

 The closest natural example in space is the Io plasma torus inside Jupiter’s magnetosphere \cite{[77.]}. Eruptive material fromIo produces a plasma torus that envelops the entirety of a moon’s orbital corridor and has been observed to remain roughly constant, even though the volcanic activity emitting the mostly heavy ions of SO$_2$, i.e., S$^+$, S$^{++}$, O$^+$, O$^{++}$ on Io is presumably sporadic \cite{[78.]}. Undoubtedly the situation for Io has major differences – the presence of the large, rapidly rotating magnetic field from Jupiter being just one factor \cite{[77.]}. 

\begin{figure}
	\centering
	     \includegraphics[width=1.0\textwidth]{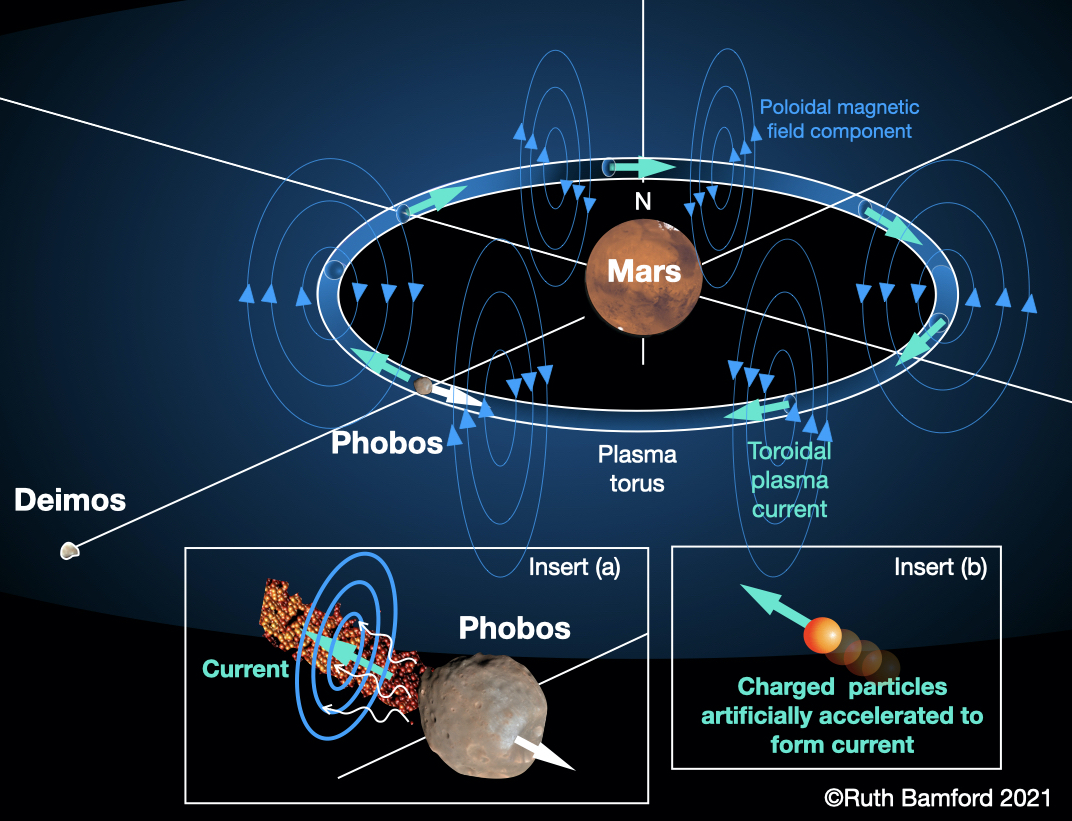}
	     \caption{Plasma released from the moon of Mars (Phobos or Deimos) (Phobos shown in insert (a)) and artificially accelerated (insert(b)) forming a plasma torus with current drive that produces a resultant magnetic field. Unlike natural systems the particles can be accelerated above the Dricer limit, to form runaway beam relatively unaffected by plasma environment and pickup processes but will need ‘kicker’ stations to ensure a closed loop is formed. The poloidal magnetic field shown created by current aids the plasma column confinement as well as forming the resultant magnetic field. If the plasma beam is mass loaded with sufficient density to allow for diffusion losses it still would take $10^{11}$ Earth years to exhaust the material of Phobos this way.}
	     \label{FIG:9}
\end{figure}
  
The most efficient source of plasma particles to form the torus is to directly evaporate them off one of the moons of Mars. This would locate the rings at their orbits. This is shown in Figure 9 for Phobos. Using the natural moons of Mars, Phobos (or Deimos), as ``plasma generators'' to ``seed'' the ring and building this plasma up to form a ring current over many months or years would minimize the resources needed. Phobos is 2.76 Mars radii above the center of Mars (about 6,000 km above the surface) and orbits in 7h and 39m whereas Deimos is in a higher orbit at 6.92$R_M$ from the centre and with a 30.3 h orbital period. Deimos would provide a large surface area to host multiple nuclear power generators and related infrastructure. Both moons orbit above the equator of Mars and Mars is tilted by a very similar angle to that of the Earth at about 25 degrees to the orbital plane of Mars around the Sun.

The moons of Mars are so small (Phobos’ diameter is 22.2 km and Deimos is 12.6 km) that they have virtually zero escape velocity so, if there was a ``plasma generator'' such as an ablation or gas release system (powered by the already referenced nuclear reactors), the plasma would easily escape from the surface. Many elements are easily ionized when exposed to solar radiation even at 1.3AU, creating equal numbers of electrons and ions of a plasma. This would naturally start to form a plasma ring as the moon moves in its orbit. After 30.3 hours (for Deimos) the moon would return to its original orbital position, and the plasma ring would start to be reinforced. Over time a permanent plasma ring could be established. Phobos being bigger, with more surface area, and closer to Mars, it returns to the starting point 4× quicker. This process would then be similar to what happens with Io, where volcanic eruptions generate the Io plasma torus inside Jupiter’s magnetosphere \cite{[77.]}. The Io-Jupiter interaction forms a complex current system accelerating particles into the poles. In the Io-Jupiter case there is an extremely strong pre-existing, natural magnetic field and very rapid rotation. This, unfortunately, will not be the same for Mars.

How quickly the plasma will drift away before Deimos or Phobos are able to get back to re-supply it at each location on the next orbit, will depend upon many factors as discussed above. The experience for Io-Jupiter is that the heavy ions stay in place longer. For Io, the torus material is composed of volcanically released sulfur, oxygen, sodium, and chlorine which then get ionized \cite{[79.]}. The composition of Phobos and Deimos is believed to be phyllosilicates or silicate minerals \cite{[80.]}. For the purposes of creating an electrical current any ions that are easily ionized are suitable. The electrons carry the current, but the opposite drift of the ions helps “anchor” the electrons to associate with the ions via charge separation electric fields. A guide toroidal field would help prevent the loss as would the current driven in the plasma loop. Once the loop is formed the current drive in the torus would result in a “pinch” effect, with a magnetic field or minor radius oriented (compressional) magnetic field component that would help inhibit outward diffusion \cite{[81.]}.

Even though Mars’ moons are not large, there would be no risk of running out of raw material for the ring itself. Using Phobos, it would take $\sim10^{11}$ Earth years to completely exhaust the available material. This is taking the total mass of Phobos of $\sim10^{16}$kg, creating a current ring of dimensions b=2000km and a $R_0=2.8R_M$ and a current of $\sim7\times10^7$Amps that requires $\sim5\times10^5$ charge carries per m$^3$ and therefore a mass of only 15kg per loop if we take a representative atomic mass of 22. This calculation is based on a composition assumed to be dominated by silicon and oxygen for Phobos, that is only singly ionized, and assuming a complete replenishment each orbit. Even with these rough numbers there is plenty of leeway.

Although a plasma torus would need to be actively driven, requiring significant levels of power, the infrastructure would be far less than any solid conductor approach as already discussed above and it offers by far the most credible approach of those we have considered here.

\section{Summary and conclusions}

If Mars is ever to be a long-term abode for human life, it will possibly need the protection of an artificially created magnetic magnetosphere of planetary dimensions. Earth’s magnetosphere helps protect the planet from the potential sterilizing effects of cosmic rays and helps retain the atmosphere from significant stripping during large solar superstorms as they pass over the planet. Here we have shown some simple calculations exploring the basic physics and engineering of what would be needed practically to create a planet sized artificial magnetic field similar to Earth’s. Clearly the resources needed would be vast. The purpose here has not been to examine the performance of a magnetic based magnetosphere at Mars, nor to justify the need for magnetic shield. Rather the intention here is to quantifiably explore the practical ways this might be done if humanity chose to do so and to make some estimate of the resources that would be involved. This is done for the first time in a scientific journal. This has deliberately been done to one significant figure precision as each approach presented would need a separate article to detail the level of technology development needed to justify more exacting figures. However, this first brush does allow a comparison of approaches and exploration of ideas.

No individual solution comes without vast technical challenges, many of which go beyond what can be described here. The primary challenge is not the intensity of magnetic field needed but the size of the required spatial dimensions. Evidence from Earth’s magnetosphere is that the magnitude of the magnetic field intensity to hold back the solar wind is about $\sim$100nT. However, to protect the whole of Mars this would need to be a continuous field over an absolute minimum area of $\sim 10^9$km$^2$ (the surface area of Mars assuming a 100km atmosphere).  To allow for such a magnetosphere to persist during the interaction with the solar wind under all conditions, this would need to be very much larger.

Of the options considered here it is unlikely that restarting Mars’ core will ever be viable option. The problem is not just the minimum of $10^{11}$, 1 Megaton hydrogen bombs needed to be distributed about the iron core to melt it, but the uncertainty that the dynamo would even restart if this was done or how long any circulation would continue, as it is currently uncertain  why Mars’ dynamo stopped in the first place - assuming Mars did once have a natural magnetic field arising from its core, like Earth. 

Solenoid loops are the next option and there are a variety of potential locations and technologies. In terms of location these range from on the planet’s surface, to stable orbits and co-incident with Mars’ moons. With an artificial system the magnitude of the magnetic field at the source and the size of the structure creating it are available to be traded. What we have shown is the advantage that a wide radius ($R_0$) solenoid loop provides not only in terms of lower magnetic field requirements at the coil surface (which would be safer to work and live around) but that the reduction in the rate of decrease of magnetic field with distance is less by $R_0^3$ making it much more efficient at covering a wider area than a small radius loop with higher field. A wide diameter current loop requires a larger physical structure to be built in space, however.

The magnetic field generating structures could be made of superconducting materials or permeant magnets both of which minimize operating power but have the disadvantage of being heavy and made from rare minerals. Alternatively, carbon nanotubes offer a potentially lighter conducting structure but are fragile and have a finite resistivity requiring continuous power and therefore power losses to overcome. 
 
We have shown that the currents required are between $\sim$0.2-0.5 GAmps in one or many solenoid loops. Whilst the power requirement will depend upon the material used, it can be estimated to be between 0.1-100 GW which is between less than one but up to 50 typical 2GW power stations. While not trivial this is not unimaginably large, especially if controlled nuclear fusion has been successfully developed as an efficient energy source in the future. 

One final approach to reduce the mass burden is to use a beamed plasma current rather than any form of solid conductor. In this scenario the current traverses the vacuum of space. To do this, the charged particles making up the current need to be accelerated to velocities where the interaction with the surround plasma environment is not sufficient to disrupt the loop of current. Exceeding the runaway limit would mean the particles would not be prone to ‘pick-up’ from solar storms. Once the magnetic field is established, the plasma current channel would reside in the relative protection of its own magnetosphere. Evidence from natural planetary observations is that even non-current driven, non-runaway plasma torii around planets (like the radiation belts) are not totally eroded by solar wind particle pick-up, though losses are inevitable. It would undoubtably be necessary to direct and replenish the plasma current via a series of aligned space stations. However, depending upon its size and location, such a relativistic particle beam could be a potential radiation hazard to transiting spacecraft. So, an alternative would be to ‘mass load’ the plasma loop by accelerating the particles to a lower non-hazardous velocity but with overwhelming much larger number density. This could be done by evaporating matter from Phoebe or Deimos, ionizing it and using electromagnetic current drive techniques to accelerate the resulting charged particles. The closest natural phenomena to this without the current drive, is the plasma torus created in Io’s orbit around Jupiter. What Io’s example shows is that a high Z plasma loop around a planet can form and persist (although in Io’s case the enormous magnetic field of Jupiter amongst other factors will help confinement). We have shown that,for Mars, the mass needed would not substantially erode the moons at approximately $\sim$15kg per orbit per loop. Using higher Z ions to form the torus will aid retention.

In conclusion, as anticipated the resources needed to create a planetary sized magnetic field are non-trivial and there is much further research to be done. What has been presented here are some unique solutions for the approaches required to create an artificial planetary sized magnetic field. 

Whilst the ideas presented here are at the scale of a planet like Mars, the principles are equally applicable to smaller scale unmagnetised objects like manned spacecraft, space stations or moon bases, creating protective `mini-magnetospheres'.

With a new era of space exploration now underway, this is the time to start thinking about these new and bold future concepts. As has been proposed by the recent White Paper for NASA’s planetary decadal survey Interdisciplinary Research in Terraforming Mars: State of the Profession and Programmatics \cite{[82.]}, there is a need to close strategic knowledge gaps and begin to further these concepts and others, in order to work toward a solution that will make colonizing Mars by humans an eventual reality. 

\section*{Acknowledgements}

This paper is dedicated to the memory of John Bradford, in thanks for many fruitful discussions. The authors would like to thank the UK Science Technology Facilities Council and RAL Space in-house research for their support and NASA. Thanks to Lasers and Plasmas Group at Instituto de Plasmas e Fusão Nuclear at Técnico Lisboa (IST), Portugal.


\end{document}